\documentclass{acmconf}
\usepackage{epsfig}
\begin{document}
\title{Simulating Resilience in Transaction-Oriented Networks}
\author{D.~Zinoviev$^{\star\dagger}$, H.~Benbrahim, G.~Meszoely$^\dagger$, and D.~Stefanescu$^{\star\dagger}$\\%
Mathematics and Computer Science Department$^\star$ and\\%
Center for Business Complexity and Global Leadership$^\dagger$, Suffolk University, Boston, MA 02114%
}
\maketitle

\vskip\baselineskip

\abstract{
The power of networks manifests itself in a highly non-linear amplification of a number of effects, and their weakness---in propagation of cascading failures. The potential systemic risk effects can be either exacerbated or mitigated, depending on the resilience characteristics of the network. The goals of this paper are to study some characteristics of network amplification and resilience. We simulate random Erd\H{o}s--R\'enyi networks and measure amplification by varying node capacity, transaction volume, and expected failure rates. We discover that network throughput scales almost quadratically with respect to the node capacity and that the effects of excessive network load and random and irreparable node faults are equivalent and almost perfectly anticorrelated. This knowledge can be used by capacity planners to determine optimal reliability requirements that maximize the optimal operational regions.
}

\section{INTRODUCTION}
The power of networks manifests itself in a highly non-linear amplification of a number of effects. Probably the most notable example is the spread of epidemics, whereby a disease can spread from one person to many through a human dynamics network~\cite{pastorsatorras2002}. The evolution of life on Earth shows, how species diversity exploded once cells started interacting with each other and sex was invented. This formed a complex network that amplified the effects of natural selection~\cite{dennett1996}. Without this amplification, natural selection would have progressed linearly at best, taking much more than four billion years to create today’s natural world. In Finance, the 2007 crisis was an exemplification on how default on a small number of mortgages brought down the likes of Lehman Brothers. At the microbiology level, RNA networks have been shown to generate complex functions that amplify metabolism, in fact creating real chemical factories~\cite{barabasi2011}. Finally, the biggest enigma is how networks of neurons amplify sensory inputs to create cognition~\cite{minsky1969}.

The flip side of amplification is network resilience (or lack of it). Systemic risk  assessment in the financial supply chain provides a situation  where the ability to evaluate the potential negative impact of blockages in the global financial supply chain may prove crucial to the well being of the banking sector and the macroeconomy. As pointed out in~\cite{haldane2009}, atomistic risk management in financial systems is unsatisfactory as it fails to give a salient risk assessment both at the nodes and for the entire system. Indeed, the effect of an idiosyncratic event at one node may propagate to other nodes, thereby resulting in financial instability or catastrophic failure of other nodes or of the entire network. Propagation of cascading failure can happen in many types of interdependent systems~\cite{buldyrev2010}. The potential systemic risk effects can be either exacerbated or mitigated, depending on the resilience characteristics of the network.

A vast body of research has been focused on the impact of the structure of the networks and nodes. Phase transitions in networks~\cite{holme2006,zapperi1997}, whereby giant components form when density reaches a critical point, are frequently used to study diffusion in networks~\cite{shi2007}. Percolation analysis has been used to measure the structural importance of particular nodes, in terrorist networks~\cite{galam2003} and soccer games~\cite{duch2010}. A number of studies have shown how the structure of a network impacts fault tolerance, showing for instance that scale free networks are particularly vulnerable to targeted attacks and immune to random attacks, whereby random networks are the opposite~\cite{albert2000,gallos2005}.

The goals of this paper are to study some characteristics of network amplification and resilience. For this study, we chose a stylized version of a network where nodes process generic transactions requiring certain capacity and processing time. We particularly focus on the relationship between the resilience of a network and the resilience of a typical node. As these services become part of a larger complex network that powers the firm's operations, the overall reliability is less than the reliability of each individual component. There is an amplification of fault.

In this study we minimize the effects of structure by using random networks~\cite{erdos1959} with homogeneous nodes. We measure amplification by varying node capacity, transaction volume, and expected failure rates.

\section{NETWORK CONFIGURATION}

In this paper, we simulate and discuss a random Erd\H{o}s--R\'enyi network~\cite{erdos1959} that consists of $N=1,600$ identical nodes representing network hosts.

Each network node represents a server that can simultaneously execute up to $C$ independent subtransactions (the nature of the subtransactions is not essential for this study). Each subtransaction takes time $\tau_0$ to complete (the time does not depend on the total load on the node). The network is simulated for the duration of $S\tau_0$.

The density of the networks is $d$, that is, of all possible $N\left(N-1\right)$ directional connections, only $dN\left(N-1\right)$ are realized. The network has no loop-back connections.

In addition to being able to execute transactions, each node can be also used for injecting transactions into the network (serve as a transaction source) and for terminating transactions, either by way of committing or aborting (serve as a transaction sink). During the simulation, transactions are injected uniformly across the network. The delays between subsequent transactions are drawn from the exponential distribution $E\left(1/r\right)$, where $r$ is the mean injection rate.

All transactions injected in the network are distributed. A master transaction $T$ consists of $L$ subtransactions $T_i$ ($i\in\{1\ldots L\}$; in our study, $L$ is drawn from the discrete normal distribution $N\left(10,4\right)$, adjusted to exclude negative values of $L$). A master transaction is committed if all its subtransactions are committed. Otherwise, the master transaction is aborted. The transaction manager is implied and not simulated.

Transactions are routed using an opportunistic routing strategy: the node for the next subtransaction is chosen uniformly at random from all neighbors of the current node. If the next node is disabled, then another neighbor is chosen. It is possible for the next subtransaction $T_{i+1}$ to be executed by the same node as the previous subtransaction $T_{i-1}$. If all neighbors are disabled, the subtransaction is aborted, and the master transaction rolls back.

Once injected in the network, a transaction has the time-to-live of $60\tau_0$. Since it takes the constant time of $1\tau_0$ for a transaction to clear a node, the fraction of transactions that are subject to aborting due to the timeout is $\approx1.2\times10^{-35}$, and this behavior may be ignored (based on the normal distribution of $L$).

We assume that distributed transactions in our network are not independent (they do not have the ACID property, which is not uncommon for distributed transactions due to the Brewer's theorem~\cite{brewer2001}). In our model, if a transaction is aborted for any reason, all other transactions that crossed path with it in the past $T$ time units ($T=10\tau_0$), are also aborted with probability $p_0=.01$.

The network nodes can become disabled in two ways. First, when a node is overloaded (the actual load at a node reaches or  exceeds its capacity $C$), it shuts down.  In real life, an overload-related shutdown may be caused by overheating, network congestion, excessive swapping or other resource constraints.

Second, the network nodes may fail randomly after an initial delay drawn from the exponential distribution $E(T_f)$. These random failures simulate the effect of the internal unreliability. Shorter time to failures correspond to less reliable nodes.

Initially, all nodes in a network are alive and can perform their tasks. Once disabled, however, a node is not restarted and remains disabled for the rest of the simulation run---recovery may not be feasible or even possible in autonomous unmanaged networks (say, sensor networks~\cite{dargie2010}). All subtransactions currently executed at a disabled node, and the corresponding master transactions, are aborted.

The network simulator has been implemented in C++ using a discrete event simulation package developed at the Mathematics and Computer Science Department of Suffolk University.

\section{SIMULATION}
\begin{figure}[tb!]\centering
\epsfig{file=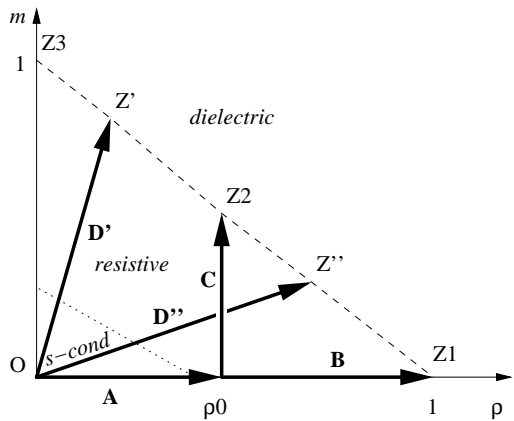,width=0.8\columnwidth}
\caption{\label{scenarios}Experimental scenarios; the dashed and the dotted lines are hypothetical phase boundaries}
\end{figure}

To study the effect of node failures on the network resilience and to propose and evaluate resilience measures, we conducted several numerical experiments, some of which are schematically presented in Figure~\ref{scenarios}.

In each experiment, the network has been simulated for a variety of combinations of node capacities and average edge densities $\left(C,d\right)$: $d\in\{$0.01, 0.011, 0.015, 0.025, 0.04, 0.055, 0.075, 0.1, 0.2, 0.3, 0.5, 0.6, 0.75, 0.85, 0.99$\}$ and $C\in\{2,3,4\ldots22\}$ (up to $C=23$ for select density values).

\subsection{Failing by Overloading}
In the first experiment, we started with a fully functional network with no injected transactions. Then we gradually increased the injection rate from 0 to $r_0$ (arrow {\bf A} in Figure~\ref{scenarios}) until at least $10^{-6}$ of all transactions would abort. Since this mode of  operation is essentially lossless, we call it {\em superconductive}. $r_0$ is defined as the maximum abort-free rate.

By injecting more than $r_0$ transactions per time unit, we partially overload the network and switch it into a {\em resistive} mode. The fraction of aborted transactions monotonically increases with the transaction injection rate $r$, until at some point the network chokes (all network nodes become overloaded and shut down) before the end of a simulation run (arrow {\bf B} in Figure~\ref{scenarios}). We denote this maximum choke-free injection rate as $r_1$, and we call this operation mode {\em dielectric}. In the same spirit, we call $r_0$ and $r_1$ phase transition injection rates.

Both $r_0$ and $r_1$ depend on the simulation running time (shorter runs allow the network to terminate choke-free for higher injection rates). However, the difference between shorter runs of $S\tau_0$ and longer runs of $2S\tau_0$ is within 5\%. All further results have been obtained for $S=84,600\tau_0$ (``one day'').

Since $r_1$ is the highest meaningful injection rate, we will sometimes normalize injection rates by introducing $\rho_0=r_0/r_1$ and $\rho=r/r_1$. We have $0\le\{\rho_0,\rho\}\le1$.

\begin{figure}[tb!]\centering
\epsfig{file=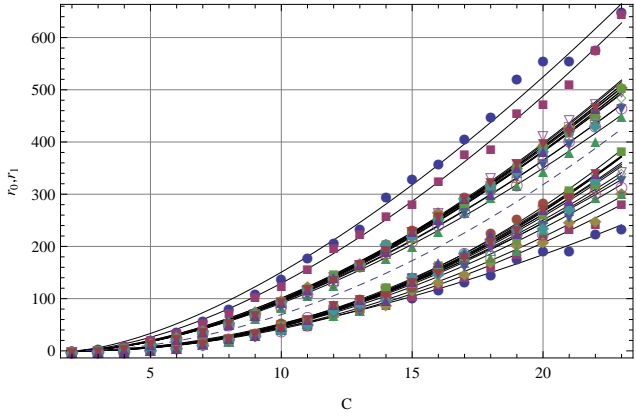,width=\columnwidth}
\caption{\label{r0r1vsC}Phase transition injection rates $r_0$ (below the dashed line) and $r_1$ (above the dashed line), in transactions per $\tau_0$, vs node capacity $C$, for various network densities $d$; solid lines represent best fit approximations}
\end{figure}

For each network configuration, we measured $r_0$ and $r_1$. Figure~\ref{r0r1vsC} shows both the experimental points and the best fit approximations that will be discussed in section~\ref{discussion}.

\subsection{Failing by Internal Faults}
In the second experiment, just like in the first one, we started with a fully functional network with no injected transactions, and gradually increased the injection rate to $r_0$ (the network is still in the superconductive state). Then, at the fixed injection rate, we started failing random nodes after random delays, simulating unrecoverable internal faults (arrow {\bf C} in Figure~\ref{scenarios}). 

\begin{figure}[tb!]\centering
\epsfig{file=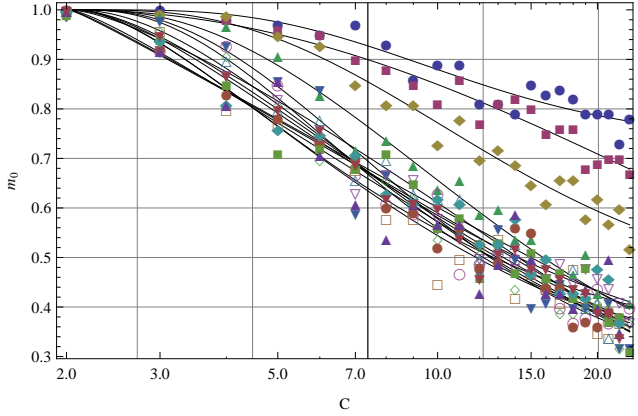,width=\columnwidth}
\caption{\label{mvsC}Phase transition node fault rate $m_0$ vs node capacity $C$, for various network densities $d$; solid lines represent best fit approximations}
\end{figure}

At the end of each simulation run, we measured the fraction of failed nodes $m$ and the state of the network (either resistive or dielectric). Let $m_0$ be the smallest $m$ that causes the network to choke and switch to the dielectric state. We call it phase transition node fault rate. Figure~\ref{mvsC} shows both the experimental values of $m_0$ and the best fit approximations. 

\subsection{Failing by Overloading and Internal Faults}
Finally, in the third experiment we combined the two mechanisms that cause network failures. 

Indeed, we learned from the previous two experiments that the phase transition between the resistive and dielectric states takes place in the points Z1 ($m=0$, $\rho=1$) and Z2 ($m=m_0$, $\rho=\rho_0$) in Figure~\ref{scenarios}. These points correspond to the first and second experiments. Point Z3 ($m=1$, $\rho=0$) is also on the phase boundary (it takes all nodes to be faulty to fail a network in the presence of zero traffic).

\begin{figure}[tb!]\centering
\epsfig{file=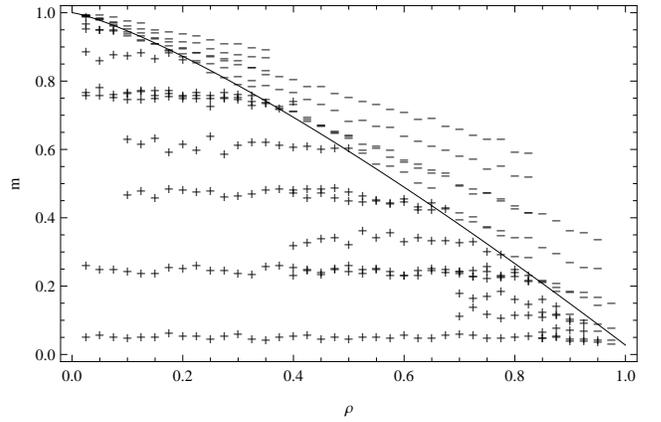,width=\columnwidth}
\caption{\label{phasespace}Phase diagram for $C=4$ and $d=0.2$. Plus ``+'' and minus ``$-$'' signs indicate resistive and dielectric points, respectively. The solid line is the best fit phase boundary}
\end{figure}

To locate the rest of the phase boundary (tentatively shown as a dashed line in Figure~\ref{scenarios}), we executed a number of simulation runs that moved the network from the original state O to various boundary states Z', Z'', etc. (arrows {\bf D'} and {\bf D''}) by simultaneously varying the injection rate and the proportion of the internally faulty nodes. The result of this experiment was a phase transition diagram for the network for each tested configuration $\left(C,d\right)$. The diagrams show the boundary between the resistive and dielectric states (we did not instrument the simulator to detect the boundary between the resistive and superconductive states, though random experiments suggest that it probably follows the dotted line in Figure~\ref{scenarios}). An example of a phase diagram for $C=4$ and $d=0.2$ is shown in Figure~\ref{phasespace}.

\subsection{Dependent Transactions}
We explored the relationship between the transaction interdependency probability $p_0$ and the simulated values of $r_0$, $r_1$, and $m_0$. Of them, only $r_1$ is statistically correlated with $p_0$: changing $p_0$ from 0 to 1 increases $r_1$ by $approx$4\%. Indeed, in the presense of strong correlation between transactions, an aborted transaction always causes a cascaded rollback, that, in turn, releaves network conjestion and allows higher injection rate---at the cost of lower commit rate.

\section{\label{discussion}DISCUSSION}

\subsection{Dense and Sparse Networks}
We observed that in all simulated scenarios, the network behavior is determined, in the first place, by the network density $d$. The borderline between different behaviors is fuzzy and lies in the range $d_0=[0.01\ldots0.02]$. In the dense networks ($d>d_0$), most performance characteristics do not depend on $d$, while in the sparse networks ($d<d_0$), the dependence on $d$ is strong to the extent that many network measures diverge as $d$ tends to 0.

\subsection{Amplification}
One goal of the study was to find the correlation between node capacity $C$ (which corresponds to the material investment into the networking infrastructure) and the aggregate network throughput expressed either in terms of $r_0$ or $r_1$. The relationship between $C$ and $r_0$ and $r_1$ for various network densities $d$ is shown in Figure~\ref{r0r1vsC}.

For both dense and sparse networks, both $r_0\left(C\right)$ and $r_1\left(C\right)$ can be approximated using a power function:
\begin{equation}
r_i\left(C\right)\approx A_i\left(C-2\right)^{\beta_i}.\label{powerlaw}
\end{equation}
The exponents $\beta_i$ for the dense networks are $\approx\!1.7$ and $\approx\!2.1$, respectively. Both $\beta_i$'s tend to 1 as $d$ tends to 0. The mantissas $A_i$ for the dense networks are $\approx\!0.7$ and $\approx\!2.8$, respectively. Both $A_i$ increase and possibly diverge as $d$ tends to 0.

We observed the quadratic amplification effect: doubling node capacity almost quadruples the throughput.

\subsection{Effect of Faulty Nodes\label{faultynodes}}
We could not easily find an explainable closed form approximation of $m_0\left(C\right)$. Eq.~\ref{erf} seems to be in good agreement with the experimental results (Figure~\ref{mvsC}).
\begin{equation}
m_0\left(C\right)\approx \frac{\left(A-1\right)\mathrm{erf}\left(\frac{\log_{10}\left(C-2\right)}\alpha - \beta\right) + \left(A + 1\right)}{2}.\label{erf}
\end{equation}

The purpose of Eq.~\ref{erf} is chiefly to estimate the dependencies between $m_0$ and $C$, not to predict them. In particular, we are not sure at this point if, as $C$ tends to infinity, all $\left(m_0\right)$s tend to 0, to a common positive asymptote or to individual positive asymptotes. Exploring this issue would require more computational resources that we can presently afford.

Based on the data that we have, we conclude that for the dense networks, the best fit curves described by Eq.~\ref{erf} converge to a value of $A$ in the range $[0..0.23]$. In other words, in the best case it would take 30\% of internally faulty nodes to fail a dense network with infinite buffer space in the presence of the highest superconductive injection rate. In the worst case, the network may fail even with negligibly few faulty nodes. In general, as the node capacity increases, the proportion of nodes that must be disabled to choke the network at the insertion rate $r_0$, decreases. This is not surprising, since $r_0$ itself scales up with $C$, so we expose the network to higher volumes of traffic.

\subsection{Equivalence of Excessive Traffic and Faulty Nodes}
\begin{figure}[tb!]\centering
\epsfig{file=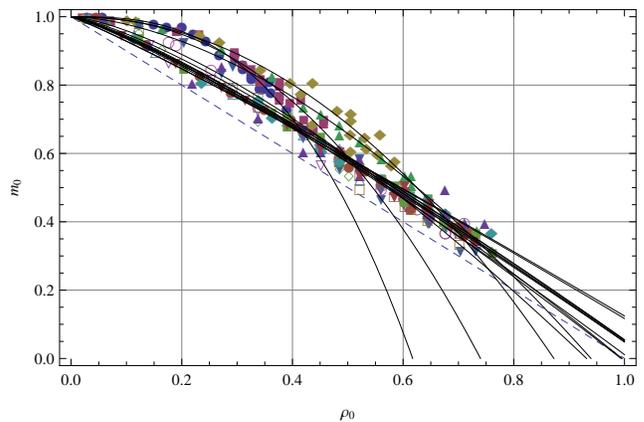,width=\columnwidth}
\caption{\label{mvsr0r1}Node fault rate $m_0$ vs network load $\rho_0$, for various network densities $d$ and capacities $C$; solid lines represent best fit approximations; the dashed line is the diagonal of the $1\times1$ rectangle}
\end{figure}

Figure~\ref{mvsr0r1} shows the node fault rate $m_0$ vs maximum superconductive injection rate $\rho_0$, for various network densities $d$ (different symbols) and capacities $C$. The two measures are closely correlated.

The solid lines represent best fit approximations (Eq.~\ref{mrho}). 

\begin{equation}
m_0\left(\rho_0\right)\approx1-A\rho_0^\beta.\label{mrho}
\end{equation}

The less dense networks correspond to the lines with the more horizontal initial segment at $\rho_0=0$.

For dense networks, the mantissa $A$ of Eq.~\ref{mrho} tends to 1, and the exponent $\beta$ tends to 1.15. For sparse networks, both parameters grow and possibly diverge as $d$ tends to 0.

To a first approximation, the relationship between the network resilience parameters $\rho_0$ and $m_0$ is almost linear, with the slope of $-1$. This means that tolerating additional superconductive traffic $\Delta\rho_0$ (that is, narrowing the gap between superconducting--resistive and resistive--dielectric injection rates) is equivalent to disabling extra network nodes $\Delta m_0$ due to internal faults, and the other way around:

\begin{equation}
\Delta\rho_0\approx-\Delta m_0.\label{equivalence}
\end{equation}

Each line in Figure~\ref{mvsr0r1} corresponds to a particular network density $d$, and the experimental points along the line correspond to various node capacities $C$. The points with higher values of $C$ have higher $\rho_0$ and lower $m_0$. If the statement about the asymptotic behavior of $m_0$ with respect to $C$ (that we made in subsection~\ref{faultynodes}) is true, than there is a convergence point $\approx\left(0.75,0.3\right)$ on the chart. No network would be able to sustain higher relative superconductive injection rate or choke with fewer faulty nodes.

\subsection{Combined External and Internal Effects}

In the first two experiments, we either exposed a healthy network to excessive traffic or faulted random nodes carrying the highest sustainable superconductive traffic. We found these mechanisms complementary and even commensurable (especially for dense networks).

In reality, a network can be simultaneously subject both to internal irreparable faults and external excessive traffic. Figure~\ref{phasespace} shows the network phase transition diagram for $C=4$ and $d=0.2$ for all possible values of $0\le m\le1$ (including $m_0$) and $0\le\rho\le1$ (including $\rho_0$). The boundary between the choking and choke-free areas was calculated using best fit parameter estimation for the Eq.~\ref{boundary}.
\begin{equation}
m_0\left(r\right)\approx1-A \rho^\beta.\label{boundary}
\end{equation}

\setlength{\textheight}{7.3in}

We found that $A\approx1$ is almost independent of either $C$ or $d$. On the contrary, $\beta$ is independent of $d$ but diminishes from 1.35 to 1.15 as $C$ increases from 2 to 9: the dependence of $m$ on $\rho$ is more linear for higher capacity networks.

Incidentally, Eq.~\ref{boundary} is identical to Eq.~\ref{mrho}, aside from the actual values of $\beta$ (which are still the same in both equations for $C\ge9$). At present, we do not know whether this is a coincidence or a rule.

\section{CONCLUSION}

Our research revealed a number of interesting characteristics of random transactional networks. We studied transaction failures as a function of two factors, random node faults and incoming transaction volume. These revealed three phases of particular interest: ``superconductive'' (no transactions fail), ``resistive'' (some transactions fail), and ``dielectric'' (all transactions fail). We found that the injection rates associated with the phase transitions, scale almost quadratically with respect to the node capacity, thus providing network throughput amplification and allowing capacity planners to determine optimal reliability requirements that maximize the superconductive region.

We also found that at the resistive-to-dielectric phase transition, the effects of excessive network load and internal, spontaneous, and irreparable node faults are equivalent and almost perfectly anticorrelated. This knowledge can be used to compensate faults in isolated unmanaged networks by properly and predictably adjusting external traffic or to determine the amount of spare nodes needed to sustain predictable bursts of traffic.

Further study is required to quantify and qualify the effects of network structure including density, modularity, and assortativity.

As discussed in the introduction, the overriding goal of this research is to study the power of networks in general.  Further study will generalize the findings to include a larger class of networks and applications whereby resilience will be substituted by capability. For example, such generalization will open new research areas in economic development whereby economic productivity is a result of complex economic networks. This research could also potentially be applied to our understanding of systemic risk and effective governance, to name a few,  through a greater appreciation of network dynamics.


\bibliographystyle{acm}
\bibliography{cs}
\end{document}